\documentclass[aps,pre,showpacs,showkeys,floatfix,twocolumn]{revtex4}

\usepackage{amsmath} \usepackage{amsfonts} \usepackage{amssymb}
\usepackage{graphicx}
\usepackage{dcolumn}
\usepackage{bm}

\begin{document}


\title{Solitons in one-dimensional nonlinear Schr\"{o}dinger lattices with a local inhomogeneity}
\author{F.\ Palmero}
\email[Corresponding author. Electronic address:]{palmero@us.es}

\thanks{Permanent address: Nonlinear Physics
Group. Escuela T\'{e}cnica Superior de Ingenier\'{\i}a
Inform\'{a}tica. Departamento de F\'{\i}sica Aplicada I. Universidad de
Sevilla. Avda.  Reina Mercedes, s/n. 41012-Sevilla (Spain)}

\author{R.\ Carretero-Gonz\'alez}
\affiliation{Nonlinear Dynamical Systems Group. Computational
Science Research Center and Department of Mathematics and
Statistics, San Diego State University, San Diego, CA 92182-7720,
USA}

\author{J.\ Cuevas}
\affiliation{Grupo de F\'{\i}sica No Lineal. Departamento de
F\'{\i}sica Aplicada I, EU Polit\'ecnica. Universidad de Sevilla,
c/ Virgen de \'{A}frica s/n, 41011-Sevilla, Spain}

\author{P.G.\ Kevrekidis}
\affiliation{Department of Mathematics and Statistics, University
of Massachusetts, Amherst, Massachusetts 01003-4515, USA}

\author{W. Kr\'olikowski}
\affiliation{Centre for Ultra-high Bandwidth Devices (CUDOS), Nonlinear Physics
Centre and Laser Physics Centre, Research School of Physical Science and
Engineering, Australian National University, Canberra, ACT 0200, Australia}

\date{\today}

\begin{abstract}
In this paper we analyze the existence, stability, dynamical formation and mobility properties of
localized solutions in a one-dimensional system described by the discrete nonlinear Schr\"{o}dinger
equation with a linear point defect. We consider both attractive and repulsive defects in a
focusing lattice. Among our main findings are: a) the destabilization of the on--site mode centered
at the defect in the repulsive case; b) the disappearance of localized modes in the vicinity of the
defect due to saddle-node bifurcations for sufficiently strong defects of either type; c) the
decrease of the amplitude formation threshold for attractive and its increase for repulsive
defects; and d) the detailed elucidation as a function of initial speed and defect strength of the
different regimes (trapping, trapping and reflection, pure reflection and pure transmission) of
interaction of a moving localized mode with the defect.
\end{abstract}



\maketitle


\section{Introduction}
\label{sec-introduction}

The past few years have witnessed an explosion of interest in
discrete models that has been summarized in a number of recent reviews
\cite{reviews}. This growth has been, to a large extent,
motivated by numerous
applications of nonlinear dynamical lattice models in areas as
broad and diverse as the nonlinear optics of waveguide arrays \cite{optics},
the dynamics of Bose-Einstein condensates in periodic potentials
\cite{bec_reviews}, micro-mechanical models of cantilever arrays
\cite{sievers}, or even simple models of the complex dynamics of the
DNA double strand \cite{peyrard}. Arguably, the most prototypical
model among the ones that emerge in these settings is the, so-called,
discrete nonlinear Schr{\"o}dinger equation (DNLS) \cite{dnls,Eil03}. DNLS
may arise as a direct model, as a tight binding approximation, or
even as an envelope wave expansion: the DNLS is one of the most
ubiquitous models in the nonlinear physics of dispersive, discrete
systems.

Perhaps the first set of experimental investigations that generated an intense interest in DNLS
type equations was in the area of nonlinear optics and, in particular, in fabricated AlGaAs
waveguide arrays \cite{7}. In the latter setting a wide range of phenomena such as discrete
diffraction, Peierls barriers (the energetic barrier that a wave needs to overcome to move over a
lattice ---see details below), diffraction management (the periodic alternation of the diffraction
coefficient) \cite{7a,7aa} and gap solitons (structures localized due to nonlinearity in the gap of
the underlying linear spectrum) \cite{7b} among others \cite{eis3} were experimentally observed.
These phenomena, in turn, led to a large increase also on the theoretical side of the number of
studies addressing such effectively discrete media.

A related area where DNLS, although it is not the prototypical model,
it still yields useful predictions both about the existence and about
the stability of nonlinear localized modes is that of optically induced
lattices in photorefractive media such as Strontium Barium Niobate
(SBN). Since the theoretical inception of such a possibility in
Ref.~\cite{efrem}, and its experimental realization in
Refs.~\cite{moti1,neshevol03,martinprl04}, there has been an
ever-expanding growth in the
area of nonlinear waves and solitons in such periodic,
predominantly two-dimensional, lattices.
A wide array of structures has been predicted and
experimentally observed in lattices induced with a self-focusing
nonlinearity, including,  e.g., discrete dipole \cite{dip},
quadrupole \cite{quad}, necklace \cite{neck} and other
multi-pulse patterns (such as e.g., soliton stripes \cite{multi}),
discrete vortices
\cite{vortex}, and
rotary solitons \cite{rings}.
Such structures have a definite potential to be used as carriers and
conduits for data transmission and processing, in the setting of all-optical
communication schemes. A recent review of this direction can be found in
Ref.~\cite{moti3} (see also Ref.~\cite{zc4}).

Finally, yet another independent and completely different physical setting
where such considerations and structures
are relevant is that of soft-condensed matter physics, where droplets
of the most recently discovered state of matter, namely of
Bose-Einstein condensates (BECs), may be trapped in an (egg-carton)
optical lattice (OL) potential produced by counter-propagating
laser beams in one, two or even all three directions \cite{bloch}.
The field of BEC has also experienced a huge growth over the past few years,
including the prediction and manifestation of modulational instabilities
(i.e., the instability of spatially uniform states towards spatially modulated
ones)
\cite{pgk}, the observation of gap solitons \cite{markus}, Landau-Zener
tunneling (tunneling between different bands of the periodic potential)
\cite{arimondo} and Bloch oscillations (for matter waves subject
to combined periodic and linear potentials) \cite{bpa_kasevich}
among many other salient features; reviews of
the theoretical and experimental findings in this area have also
recently appeared in Refs.~\cite{konotop,markus2}.

While DNLS combines two important features of many physical lattice
systems, namely nonlinearity and periodicity, yet another element
which is often physically relevant and rather ubiquitous is disorder.
Localized impurities are well-known in a variety of settings to
introduce not only interesting wave scattering phenomena \cite{marad},
 but also to
create the possibility for the excitation of impurity modes,
which are spatially localized oscillatory states at the impurity
sites \cite{Lif}. Physical applications of such phenomena arise, e.g.,
in  superconductors \cite{andreev}, in the dynamics of the
electron-phonon interactions \cite{tsironis}, in the
propagation of light in dielectric super-lattices with embedded
defect layers \cite{soukoulis} or  in defect modes arising in photonic
crystals \cite{Joann}.

In the context of the DNLS, there have been a number of interesting
studies in connection to the interplay of the localized modes with
impurities. Some of the initial works were either at a quasi-continuum
limit (where a variational approximation could also be implemented
to examine this interplay) \cite{forinash} or at a more discrete level
but with an impurity in the coupling \cite{Kro_Kiv96} (see also in
the latter setting the more recent studies of a waveguide bend
\cite{takeno,agrotis} and the boundary defect case of Ref.~\cite{longhi}).
More recently the experimental investigations of Refs.~\cite{Pes_et99,Mor_et02}
motivated the examinations of linear \cite{molina} and nonlinear
\cite{molina,KKK03} defects in a DNLS context. In the photorefractive
context, further recent experimental work has illustrated blocking
effects to a probe beam from either bright or dark soliton beams in
defocusing waveguide arrays \cite{kipp}.

Our aim in the present work is to systematically examine the properties of the focusing DNLS
equation in the presence of both an attractive and a repulsive linear impurity. Our first aim is to
present the full bifurcation diagram of the localized modes in the presence of the impurity and how
it is {\em drastically} modified in comparison to the case of the homogeneous lattice. The relevant
bifurcations are quantified whenever possible even analytically, in good agreement with our full
numerical computations. A second problem that is examined is that of the threshold for the
formation of solitary waves and how it is {\em systematically} affected by the presence of
impurities both in the repulsive and in the attractive case. This is motivated by the recent
examination of the relevant threshold in the homogeneous lattice \cite{Kpre} and its connection
with experiments in focusing \cite{mora} (and even defocusing \cite{rosberg}) waveguide arrays.
Finally, in the same spirit as that of Ref.~\cite{molina}, but for attractive and repulsive
impurities, we systematically investigate the interaction of an incoming solitary wave with the
localized impurity, identifying the main observed regimes as being trapping, reflection with
trapping, pure reflection and pure transmission.


This paper is organized as follows. In section II, we introduce
the model. In Section III, we analyze the existence and
stability of localized excitations in a system described by the
DNLS with the linear impurity. In Section IV, we
examine the (energy/initial amplitude) threshold for soliton formation.
In Section V
we present our results related to the interaction of moving
localized excitations with the impurity and, finally, in Section
VI, we summarize our findings and present our conclusions.

\section{The model}
We consider a discrete system with a defect described by the DNLS
equation as
\begin{equation}
\label{DNLS}
 i \dot{\psi}_n + \gamma |\psi_n|^2 \psi_n+C
(\psi_{n+1}+\psi_{n-1}) + \alpha_n \psi_n =0,
\end{equation}
where $\psi_n$ is the complex field at site $n$ ($n=1 \ldots N$); $\gamma$ is the anharmonicity
parameter, $C$ the coupling constant and parameters $\alpha_n$ allow for the existence of local,
linear inhomogeneities. In this paper, we  consider a single point defect, thus $\alpha_n=\alpha
\delta_{n,n_0}$, that can be positive (attractive impurity)  or negative (repulsive impurity). In
general, the presence of an on--site defect would affect the nearest neighbor coupling, and
Eq.~(\ref{DNLS}) should be modified to take this effect into account, as in Ref. \cite{Led03}. On
the other hand, this inhomogeneity in the coupling can be avoided using different techniques, for
example, in nonlinear waveguide arrays, changing slightly the separation between defect waveguide
and its nearest neighbors, as it has been done in Ref. \cite{7a}. In this work, we will assume that
the coupling parameter $C$ is independent on the site and positive.

Upon renormalization of parameters, we consider $\gamma=1$ (focusing case). Note that the
defocusing case ($\gamma<0$) can be reduced, under the staggering transformation
$\psi_n\longrightarrow(-1)^n \psi_n$, to the previous one with opposite sign of the impurity
$\alpha$. Also, under the transformation $\psi_n \longrightarrow \psi_n e^{2 i C t}$,
Eq.~(\ref{DNLS}) can be written in the standard form
\begin{equation}
\label{DNLS2}
 i \dot{\psi}_n + \gamma |\psi_n|^2 \psi_n+C \Delta \psi_n + \alpha_n \psi_n
 =0,
\end{equation}
where $\Delta \psi_n=\psi_ {n+1}+\psi_{n-1}-2\psi_n $ is the
discrete Laplacian. Throughout this work, we use the form given by
Eq.~(\ref{DNLS}).

The DNLS (\ref{DNLS}) conserves two dynamical invariants, the Hamiltonian

\begin{equation}\label{ham}
H=-\sum_n \frac{1}{2}|\psi_n|^4+C (\psi^*_n \psi_{n+1}+\psi^*_n
\psi_{n-1})+\alpha_n |\psi_n|^2,
\end{equation}
with canonical variables $q_n=\psi_n$ and $p_n=i \psi_n^*$, and
the (squared $L^2$) norm or optical power
\begin{equation}
P=\sum_n |\psi_n|^2.
\end{equation}

\section{Stationary solutions}

In order to study solitons in the system described by
Eq.~(\ref{DNLS}), we aim to look for stationary solutions with
frequency $\omega$. Thus, substituting
\begin{equation}\label{stat}
\phi_n=e^{i\omega t} \varphi_n,
\end{equation}
The stationary analog of Eq.~(\ref{DNLS}) then reads
\begin{equation}
\label{SDNLS}
 -\omega \phi_n +C (\phi_{n+1}+\phi_{n-1}) + \phi_n^3 + \alpha_n
\phi_n=0.
\end{equation}

Some of the properties of solitons are related to the existence
(or not) and properties of linear localized modes. These modes
arise when an inhomogeneity appears and can be obtained from
the linearized form (around the trivial solution $\phi_n=0,
\forall$ $n$) of Eq.~(\ref{SDNLS}). In this case, and considering
an inhomogeneity located at the first site of the chain and with periodic
boundary conditions, the problem reduces to solving the eigenvalue
problem
\begin{eqnarray}
\label{lsystem}
 \left[
\begin{array}{cccccc}
\alpha & C & 0 & . & . & C \\
C & 0 & C & 0& . & 0 \\
0 & C & 0 & C & . & . \\
. & . & . & . & . & . \\
. & . & . & C & 0 & C \\
C & . & . & . & C & 0 \end{array} \right] & \left[
\begin{array}{c}  \phi_0  \\
\phi_1 \\  . \\ . \\ \phi_{N-2}
\\\phi_{N-1}
\end{array} \right] = \omega  \left[
\begin{array}{c}  \phi_0  \\
\phi_1 \\  . \\ . \\ \phi_{N-2}
\\\phi_{N-1}
\end{array} \right],
\end{eqnarray}
that is a particular case of the eigenvalue problem studied in Ref.~\cite{PDER05}. There it was
shown that, if $\alpha \neq 0$, the solution corresponds to  $N-1$ extended modes and an impurity
localized mode. Also, if $N$ becomes large, the frequencies of extended modes are densely
distributed in the interval $\Omega \in [-2 C, 2C]$ and the localized mode can be approximated by

\begin{figure}
\begin{center}
\includegraphics[scale=0.5]{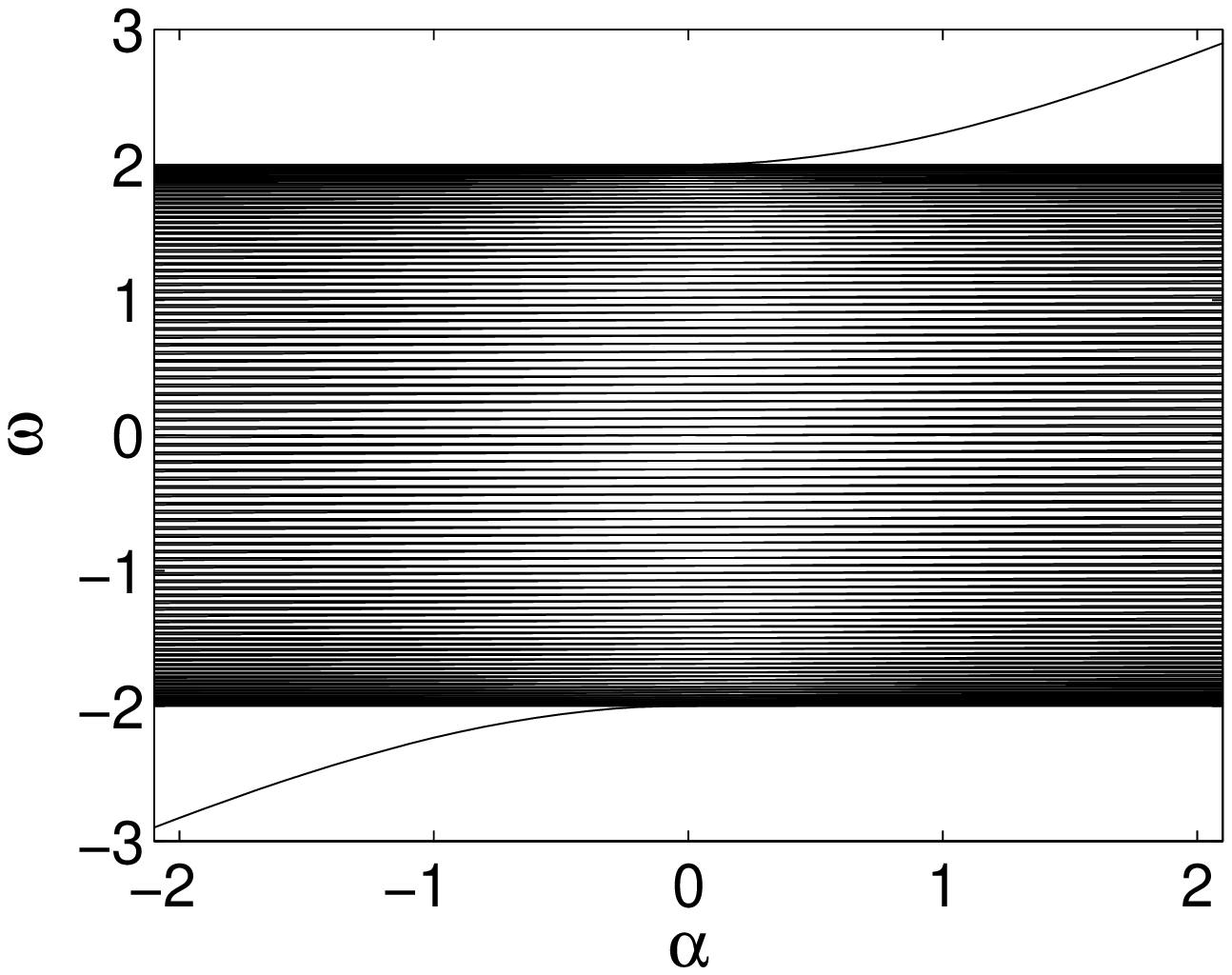}\\[1.0ex]
\includegraphics[scale=0.5]{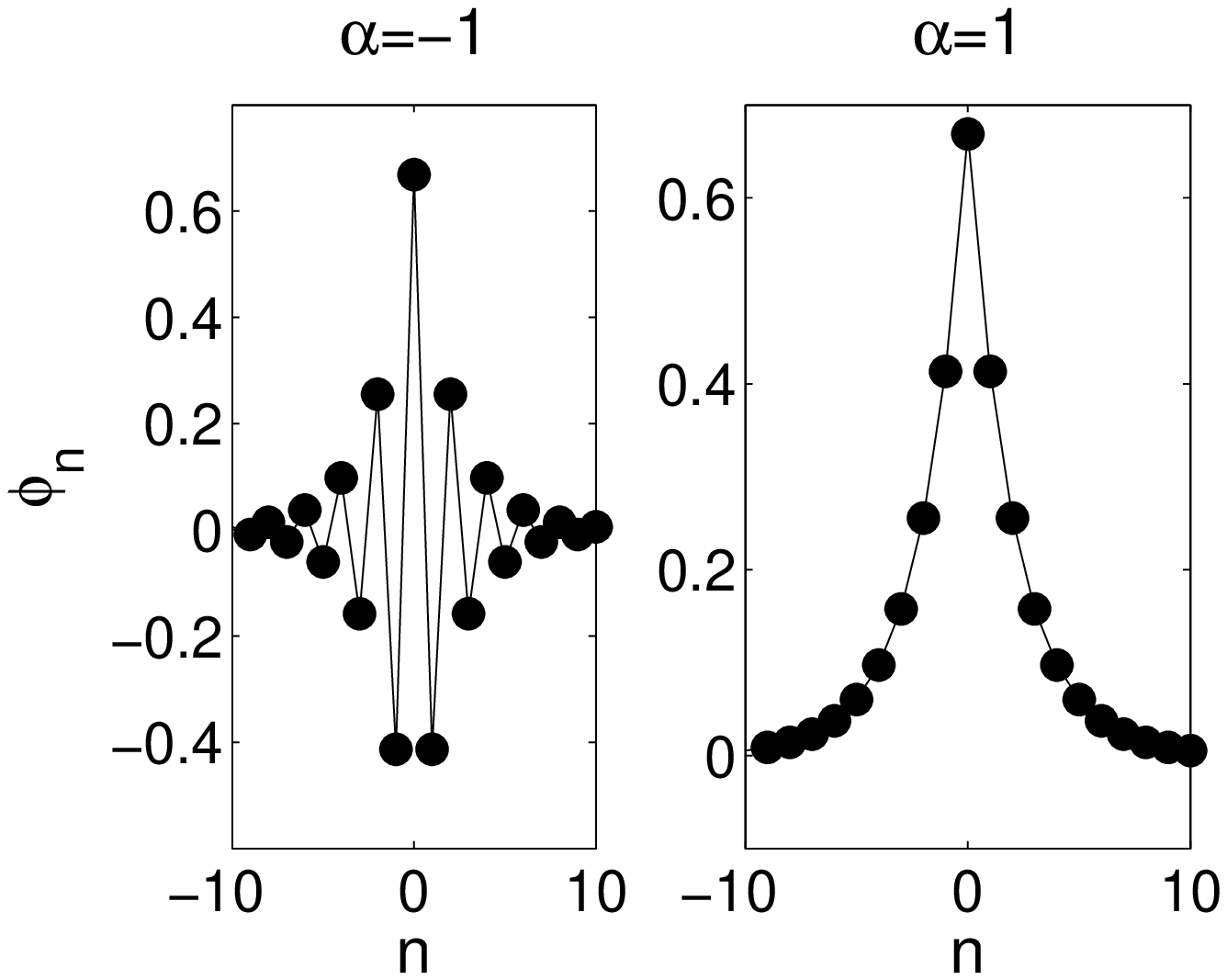}
\end{center}
\caption{Linear modes: the top panel shows the dispersion
relation as function of impurity parameter $\alpha$ (notice the
impurity mode outside of the interval $[-2 C, 2 C]$). The linear modes
are normalized ($\sum_n|\phi_n|^2=1$), the impurity is located at
$n=0$, and periodic boundary conditions are considered.
The bottom panels depict examples of the profiles of the
impurity modes.
Bottom-left panel: profile for $\alpha=-1$ (repulsive impurity),
and bottom-right panel  profile for $\alpha= 1$ (attractive impurity).
In all cases $N=200$ and $C=1$.}
\label{lmodes}
\end{figure}

\noindent\underline{$\alpha>0$ (attractive impurity):}
\begin{equation}
\phi_n=\phi_0\left[\left(\frac{\alpha}{2C}+
\beta
\right)^{-\!n}\!\!+
\left(\frac{\alpha}{2C}+
\beta
\right)^{n-N}\right],
\end{equation}
and
\begin{equation}
\omega=2 C \left|
\beta
\right|,
\quad \beta \equiv \sqrt{\frac{\alpha^2}{4C^2}+1}
\end{equation}
with an in-phase pattern (see bottom-right panel in Fig.~\ref{lmodes}).

\noindent\underline{$\alpha<0$ (repulsive impurity):}
\begin{equation}
\phi_n=(-1)^n\phi_0\left[\left(\frac{\alpha}{2C}+
\beta
\right)^{\!-n}\!\!+(-1)^N
\left(\frac{\alpha}{2C}+
\beta
\right)^{n-N}\right],
\end{equation}
and
\begin{equation}
\omega=-2 C \left|
\beta
\right|,
\quad \beta \equiv \sqrt{\frac{\alpha^2}{4C^2}+1}
\end{equation}
with a staggered pattern (see bottom-left panel in Fig.~\ref{lmodes}).
In both cases $\phi_0$ is an arbitrary constant. In Fig.~\ref{lmodes}
we depict the linear mode spectrum as a function of the inhomogeneity
parameter $\alpha$ (top panel) and examples of the profiles of
the ensuing localized modes (bottom panels).

In order to explore  the existence and stability of the nonlinear
stationary states described by Eq.~(\ref{SDNLS}), we have used the
well-known technique based on the concept of continuation from the
anti-continuum (AC) limit using a Newton-Raphson fixed point algorithm
\cite{MA96}. Also, a standard linear stability analysis of
these stationary states has been performed, using the ansatz
\begin{equation}
\phi_n=[\phi_{\rm sol}+\epsilon(a_n \exp(\lambda t)+ b_n
\exp(\lambda^* t)] \exp(i \omega t),
\end{equation}
and solving the ensuing eigenvalue problem.
$\phi_{\rm sol}$ is the solution of Eq.~(\ref{SDNLS})
with frequency $\omega$, $\lambda$ is the linearization
eigenvalue and $\lambda^*$ its complex conjugate. Due to
symmetries of the system, the  eigenvalues appear in quartets
(if $\lambda$ is an eigenvalue, so are $\lambda^{*}$,
$-\lambda$ and $-\lambda^{*}$). Furthermore, the U$(1)$
invariance of the equation (the so-called phase or
gauge invariance) leads to the  existence of a pair
of zero eigenvalues. If the remaining eigenvalues are
imaginary, the state is linearly stable and, on the contrary, the
presence of a eigenvalue with a nonzero real part implies
instability.

In the homogeneous lattice case of $\alpha=0$, fundamental stationary modes are
well known to exist and be centered either
on a lattice site or between two adjacent lattice sites \cite{dnls}.
The site-centered solitary waves are always stable, while the
inter-site centered ones are always unstable \cite{dnls}.

\begin{figure}
\begin{center}
\includegraphics[scale=0.5]{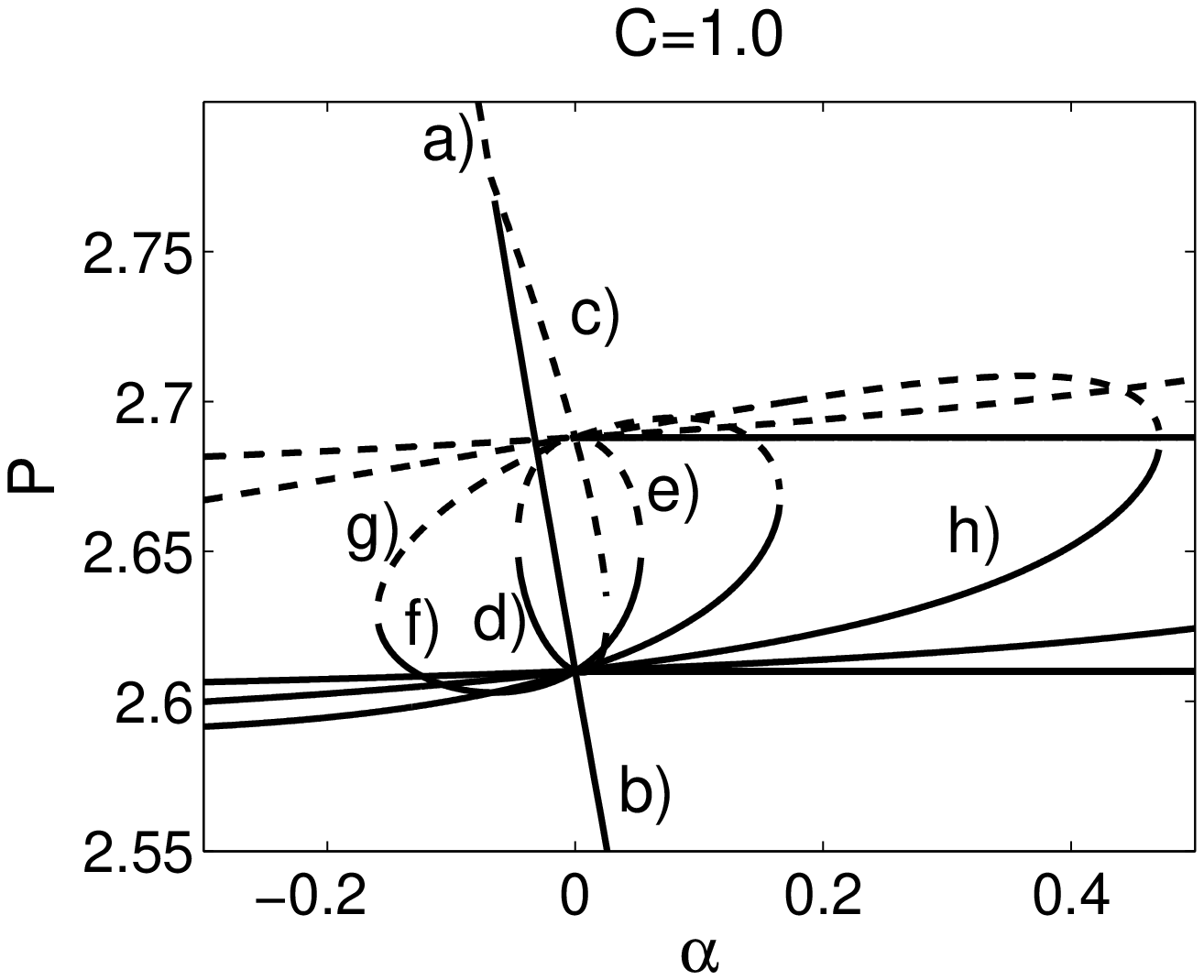}\\[2.0ex]
\includegraphics[scale=0.5]{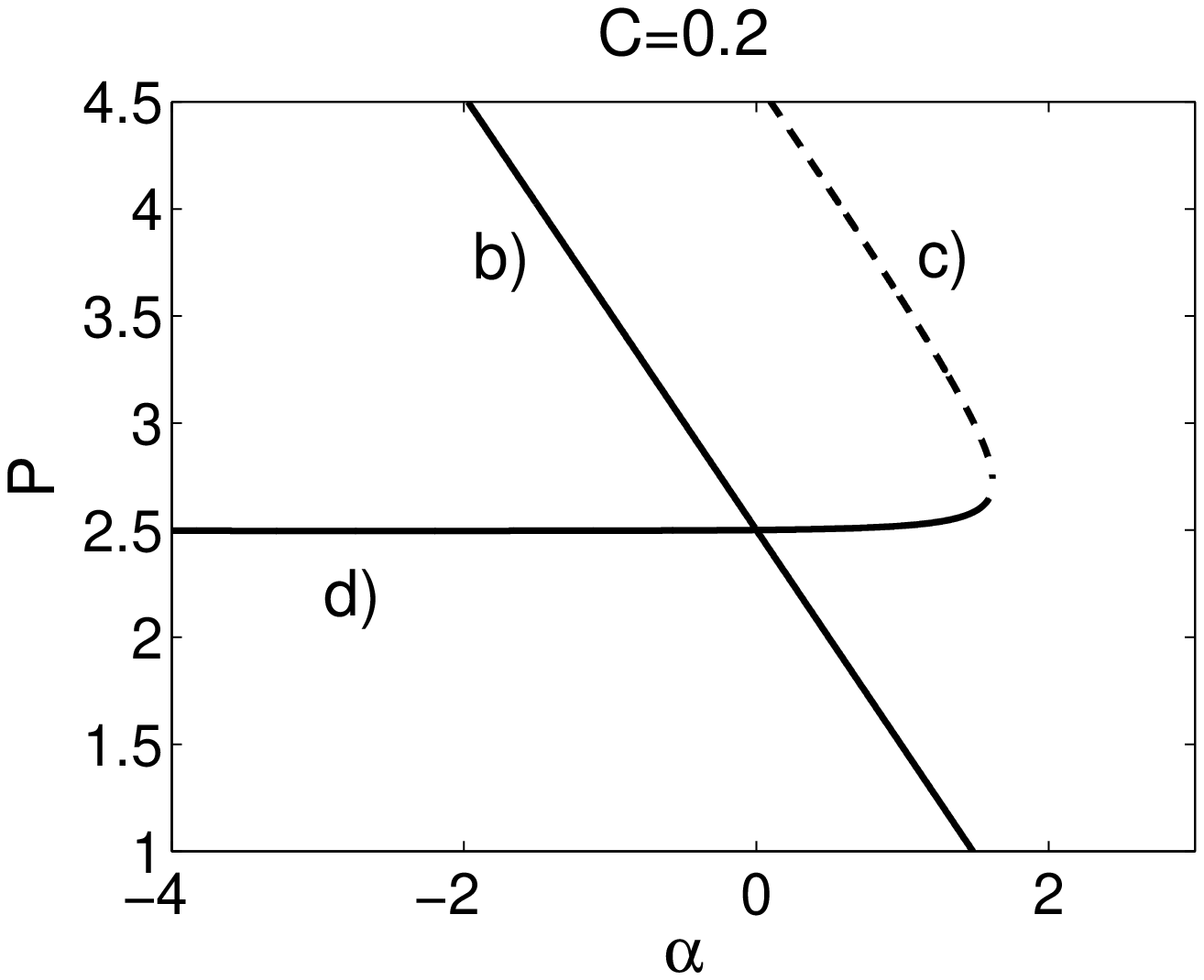}
\end{center}
\caption{Bifurcation diagram of stable (solid line) and unstable
(dashed line) nonlinear modes. Shown is the power $P$ as a function
of the impurity parameter $\alpha$. In all cases $N=100$ and
$\omega=2.5$. The top panel is for $C=1.0$, while the bottom one is for
$C=0.2$. The branch designation is as follows: a) Unstable soliton
centered at the impurity ($n=n_0$), b) stable on-site soliton
centered at $n=n_0$, c) Unstable inter-site soliton centered at
$n=n_0+0.5$, d) stable on-site soliton at $n=n_0+1$, e) unstable
inter-site soliton at $n=n_0+1.5$, f) stable on-site soliton at
$n=n_0+2$, g) unstable inter-site soliton at $n=n_0+2.5$, and h)
stable on-site soliton at $n=n_0+3$. The stable on-site mode
located at the impurity, in the homogeneous case, disappears for a
coupling value of $C\simeq 1.25$ due to resonances with the phonon
band.} \label{bif}
\end{figure}

In order to study the effects of the inhomogeneity on the
existence and properties of localized modes, we have performed a
continuation from the homogeneous lattice case of $\alpha=0$.
We found that, if
$\alpha$ increases, ($\alpha>0$, attractive impurity case), the
amplitude of the stable on-site mode decreases, while if $\alpha$
decreases ($\alpha<0$, repulsive impurity case), in general, the
stable on-site soliton localized at the impurity merges with the
unstable inter-site centered one localized between impurity and its
neighboring site (beyond some critical value of $|\alpha|$)
and the resulting state becomes unstable. Notice that, at heart,
the latter effect is a pitchfork bifurcation as the on-site mode collides
with both the inter-site mode centered to its right, as well as with
the one centered to its left.

In Fig.~\ref{bif} we show a typical bifurcation scenario where,
for a fixed value of the frequency $\omega$ and the coupling
parameter $C$, we depict the mode power $P$ corresponding to
different on-site and inter-site localized modes as a function of
impurity parameter $\alpha$. If we denote as $n_0$ the site of
the impurity, when $\alpha>0$ increases, we found that the
unstable intersite soliton localized at $n=n_0+0.5$ disappears in
a saddle-node bifurcation with the stable site soliton localized
at $n=n_0+1$. Also, if we continue this stable mode, when $\alpha$
decreases, and for a given value $\alpha=\alpha_c<0$, it also
disappears together with the unstable mode localized at $n=n_0+1.5$ through
a saddle-node bifurcation. If we increase again the impurity
parameter, this unstable mode localized at $n=n_0+1.5$ bifurcates
with the stable site mode localized at $n=n_0+2$ for a critical
value of parameter $\alpha=\alpha_c'>0$ through a saddle-node
bifurcation again, and it could be possible to continue this
bifurcation pattern until a site $n_0+k$, where the value of site
$k$ increases with the value of the coupling $C$ and the frequency
$\omega$ parameters. This scenario is similar to the one found in
previous studies with different kinds of impurities \cite{takeno,
KKK03} and appears to be quite general. It should be noted that
when the coupling parameter increases, more
bifurcations take place, in a narrower interval of power $P$ and
impurity parameter $\alpha$ values.

Some of the particularly interesting experimentally tractable
suggestions that this bifurcation picture brings forward are
the following:
\begin{itemize}
\item A localized mode centered at the impurity may be impossible
for sufficiently large attractive impurities (because the amplitude
of the mode may decrease to zero), while it may be impossible to
observe also in the defocusing case due to the instability induced
by the pitchfork bifurcation with its neighboring inter-site configurations.
\item A localized on--site mode centered at the neighborhood of
the impurity should not be possible to localize for sufficiently
large impurity strength both in the attractive {\em and} in the
repulsive impurity case.
\end{itemize}

\begin{figure}
\begin{center}
\includegraphics[scale=0.5]{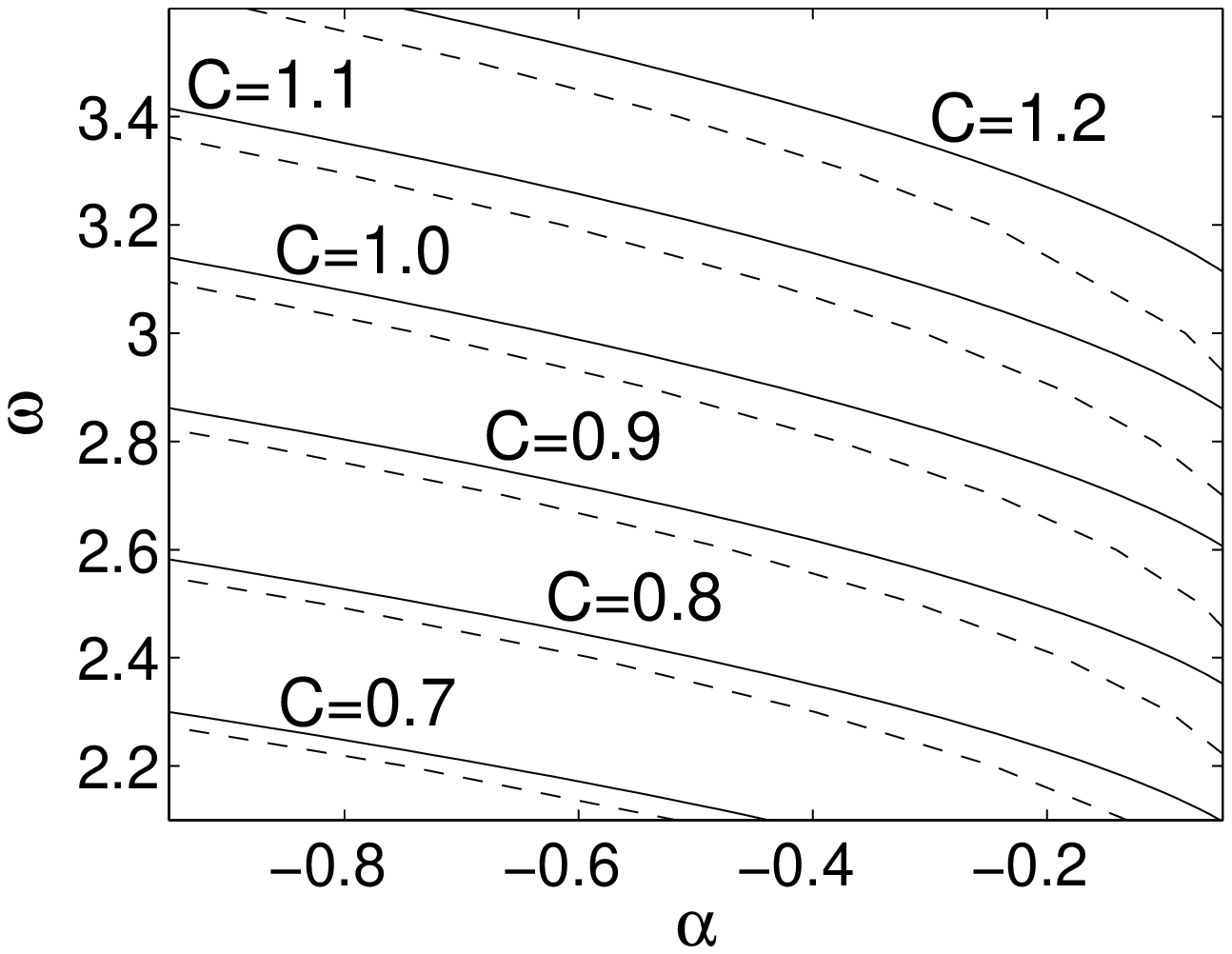}\\[1.0ex]
\includegraphics[scale=0.5]{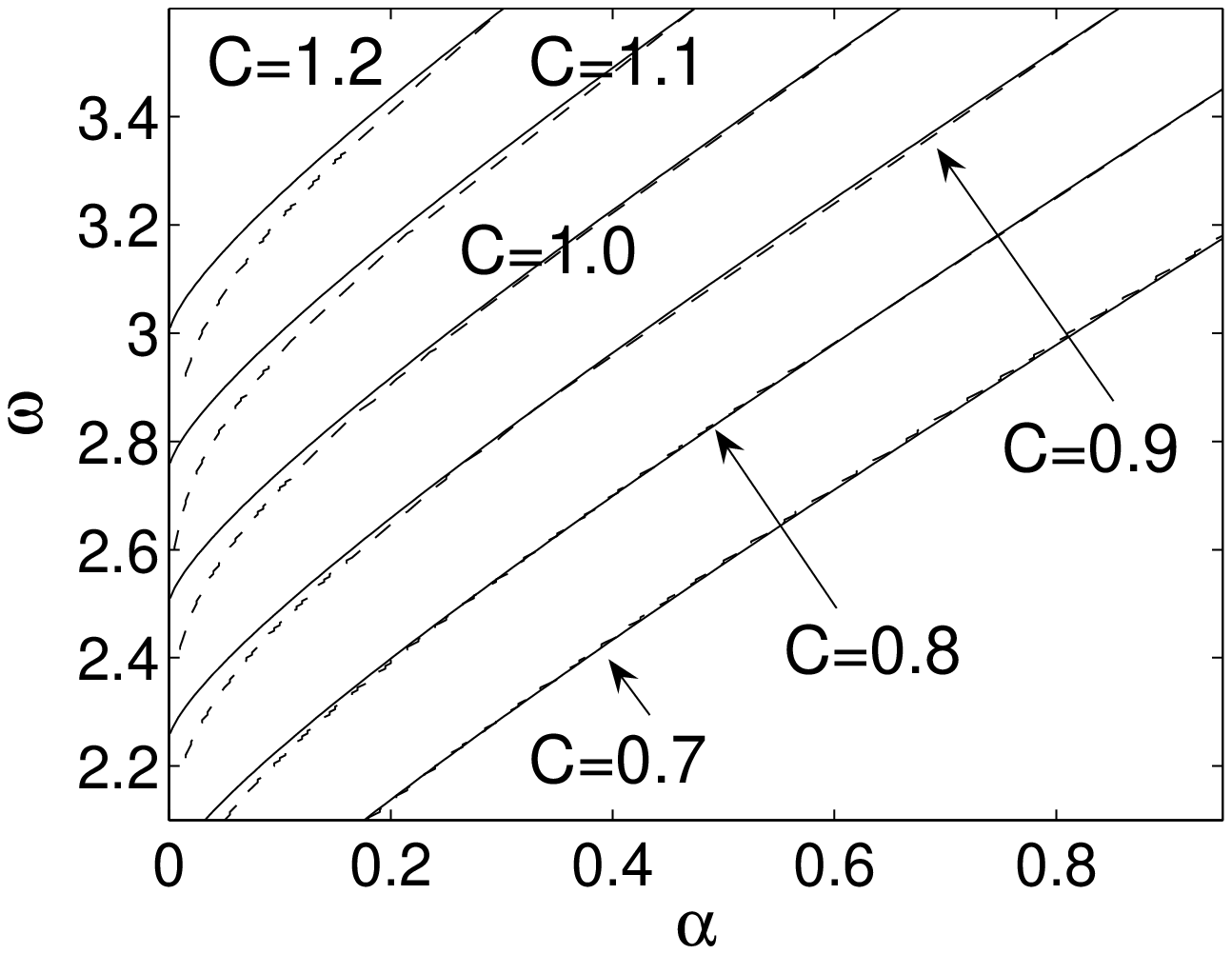}
\end{center}
\caption{Bifurcation loci corresponding to the bifurcation between
the on-site localized mode at the impurity ($n=n_0$) and its
neighbor inter-site breather ($n=n_0+0.5$) (top panel), and to the
bifurcation between the on-site localized mode next to the
impurity ($n=n_0+1$) and its neighbor inter-site breather
($n=n_0+0.5$) (bottom panel),  for different values of parameter $C$.
Dashed lines correspond to numerical results and continuous lines
to approximate analytical calculations.} \label{bif_point}
\end{figure}

We have also performed a more detailed study of the bifurcation
between the on-site nonlinear mode centered at the impurity and
its inter-site and one-site neighbor. Thus, we have determined
that, for a given value of the coupling parameter $C$, the
corresponding critical value of impurity  parameter
$\alpha=\alpha_c$. Note that this bifurcation takes place only if
$\alpha$ is negative (repulsive impurity). In case of $\alpha$
positive (attractive impurity), the inter-site solution disappears
in a saddle-node bifurcation with the on-site wave centered at the
site next to the impurity. In these cases, via an analysis of
invariant manifolds of the DNLS map, and following the method
developed in Ref.~\cite{Gui07} (see Appendix A), some approximate
analytical expressions corresponding to this bifurcation point can
be obtained. Fig.~\ref{bif_point} shows the comparison between the
exact numerical and the approximate analytical results. In
general, for a fixed value of the coupling parameter $C$, the
critical value of the frequency increases with $|\alpha|$.

\section{Threshold for Solitary Wave Formation}

We now examine the problem of solitary wave {\em formation}, i.e., whether there exists a minimal,
say, amplitude threshold for a compactum of initial data $u_n(0)=A \delta_{n,k}$ to nucleate a
localized mode. The recent work of Ref.~\cite{Kpre} suggests that a good approximation to the
amplitude of a single-site initial condition at site $k$ required to nucleate a nonlinear localized
mode at that site is given by

\begin{equation}
\label{Panos_het} - \frac{A^4}{2}+(2 C-\alpha_{k}) A^2<0.
\end{equation}
In this expression, $\alpha_{k}$ is the impurity parameter value at site $k$,
and $A$ the amplitude of the initial condition \cite{Kpre}.

\begin{figure}
\begin{center}
\includegraphics[width=8.4cm]{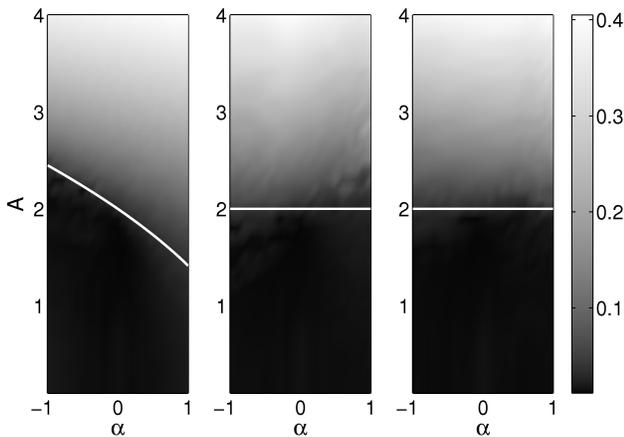}
\end{center}
\caption{Localization as function of amplitude $A$ and impurity
parameter $\alpha$ for a single excitation $\varphi_n(t=0)=A
\delta_{n,k}$. The left panel corresponds to the excitation
localized at the impurity ($k=n_0$), the center panel to the
excitation localized at the first neighbor of the impurity
($k=n_0+1$), while the right panel to the second neighbor of the
impurity ($k=n_0+2$). The solid line depicts in each case the theoretical
threshold given by Eq.(\ref{Panos_het}). In all cases $N=200$ and
$C=1.0$.}
\label{nucle}
\end{figure}

In order to study the effect of the impurity on this magnitude, we
have performed numerical simulations initially ``seeding'' energy at different
sites of the lattice (either at the impurity or at its neighbors).
After a transient state, we have analyzed the existence of localized modes
on the chain. To measure the localization of a state we have
introduced the localization of an initial
excitation of amplitude $A$, $L(A)$,
as
\begin{equation}
L(A)=\frac{\sum_n |\varphi_n|^2}{(\sum_n|\varphi_n|)^2}.
\end{equation}
Thus, for a single excited particle we have $L=1$, and if we have
$n$ excited particles (with the same amplitude, and the rest with
zero amplitude), $L=1/n$. In general, $1/N \leq L \leq 1$.

In Fig.~\ref{nucle} we summarize our numerical results and
analytical prediction. In general, when a single perturbation is
located on the impurity, numerical and analytical results are in good
agreement (left panel of Fig.~\ref{nucle}).
On the other hand, when the perturbation is located in
other (nearby to the impurity) sites of the chain,
the excitations of impurity dynamics play a significant role,
and numerical and analytical thresholds show a
slight divergence (middle panel of Fig.~\ref{nucle}).
However, when the perturbation is located far enough of
the impurity (that for the purposes of formation of a localized mode,
we return to the limit of a ``homogeneous'' lattice),
the effect of the impurity is negligible, and the
threshold  corresponding to homogeneous case is in good agreement
with the numerical data, as can be appreciated in the
right panel of Fig.~\ref{nucle}.

These results also suggest an immediately testable experimental
prediction, namely that thresholds such as the ones reported
in Ref.~\cite{mora} (see also Ref.~\cite{rosberg} for the defocusing case)
should be directly affected by the presence of a localized impurity.
In particular, an attractive linear impurity facilitates the formation
of localized modes, by decreasing the threshold of their formation,
while the opposite is true for repulsive impurities that increase the
corresponding threshold.

\section{Interaction of a moving localized mode with a single
impurity}

Early studies of the DNLS had shown that discrete solitary
waves in the DNLS can propagate along
the lattice with a relatively small loss of energy \cite{Eil86}, and
more recent work suggests that such propagating solutions
might exist, at least for some range of control parameters
\cite{Fed91,Eil03,Go04}; nevertheless, genuinely traveling
solutions are not present in the DNLS, but only in variants
of that model (such as the ones with
saturable or cubic-quintic nonlinearity)
\cite{papers0607}.

In this section we study the
interaction of propagating (with only weak radiative losses) localized
modes with the impurity. Thus, we  consider a nonlinear
localized mode, far enough from the impurity, of frequency $\omega$,
and perturb it by adding a thrust $q$ to a stationary breather
$\phi_n$ \cite{Cue06} , so that:
\begin{equation}\label{thrust}
\varphi_n(t=0) = \phi_n e^{iqn}.
\end{equation}
This is similar in spirit to the examination of Ref.~\cite{molina},
although we presently examine both attractive and repulsive impurities.
In the remainder of this study we consider $\omega=2.5$ and $C=1$,
but we have checked that a similar scenario emerges for other
values of the frequency $\omega$.

\begin{figure}
\begin{center}
\includegraphics[width=8.8cm]{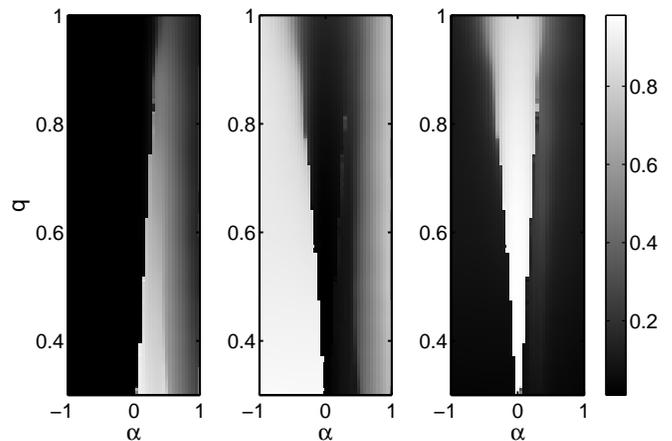}
\end{center}
\caption{Power trapping (left), reflection (center) and
transmission (right) coefficients as function of impurity
parameter $\alpha$ and initial thrust $q$. In all cases $N=1000$
and $C=1$.}\label{coef}
\end{figure}

In general, if $q$ is large enough, the soliton moves with a small
loss of radiation. We have calculated, as a function of parameters
$q$ and $\alpha$, the power and energy that remains trapped by the
impurity, reflected and transmitted along the chain, and determined
the corresponding coefficients of trapping, reflection and
transmission, defined as the fraction of power (energy) that is
trapped, reflected or transmitted. In Fig.~\ref{coef}  we summarize
our results.



We can essentially distinguish four fundamental regimes:
\begin{itemize}
\item[(a)] Trapping. If the parameters $q$ and $\alpha$ are small enough,
and the impurity is attractive, nearly all the energy remains
trapped at the impurity, and only a small fraction of energy is
lost by means of phonon radiation. An example of this phenomenon
is shown in Fig.~\ref{Fourier} (top). In this case, the central
power (power around the impurity) before the collision is nearly
zero. When the localized mode reaches the impurity, it loses power
as phonon radiation and remains trapped. The analysis of the
Fourier spectrum of this trapped breather, carried out after the
initial decay  and at an early stage of the evolution, shows a
frequency close to the initial soliton frequency, as shown in
Fig.~\ref{Fourier} (bottom). We have observed that, in general, this
frequency is  slightly smaller than that of the incident soliton,
and, in consequence, it has even smaller energy (in absolute
value) and power than the corresponding nonlinear mode with the
frequency of incident soliton.

In this particular case, corresponding to $q=0.3$ and
$\alpha=0.2$, the initial incident wave (after perturbation)
has power $P=2.61$ and energy $E=-5.40$ and the stationary mode,
trapped at the impurity, with the same frequency, has $P=2.17$
and $E=-4.73$. Thus, the incident breather can activate this
nonlinear mode, and nearly all energy and norm remains trapped. In
all simulations we have detected similar phenomena, as reported
recently in a Klein-Gordon system \cite{Cue02, Alv06}.

\begin{figure}
\begin{center}
\includegraphics[scale=0.5]{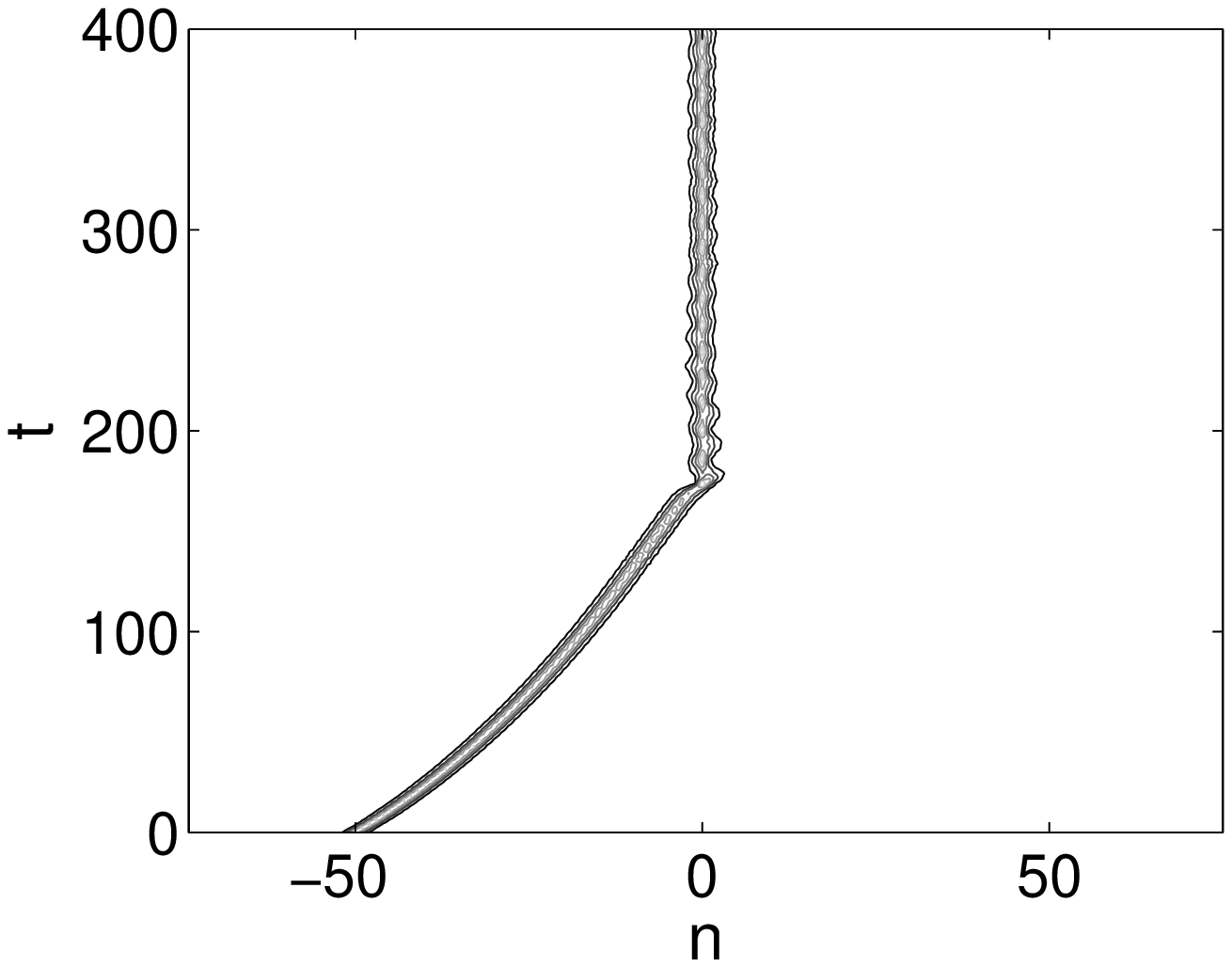}\\[1.0ex]
\includegraphics[scale=0.5]{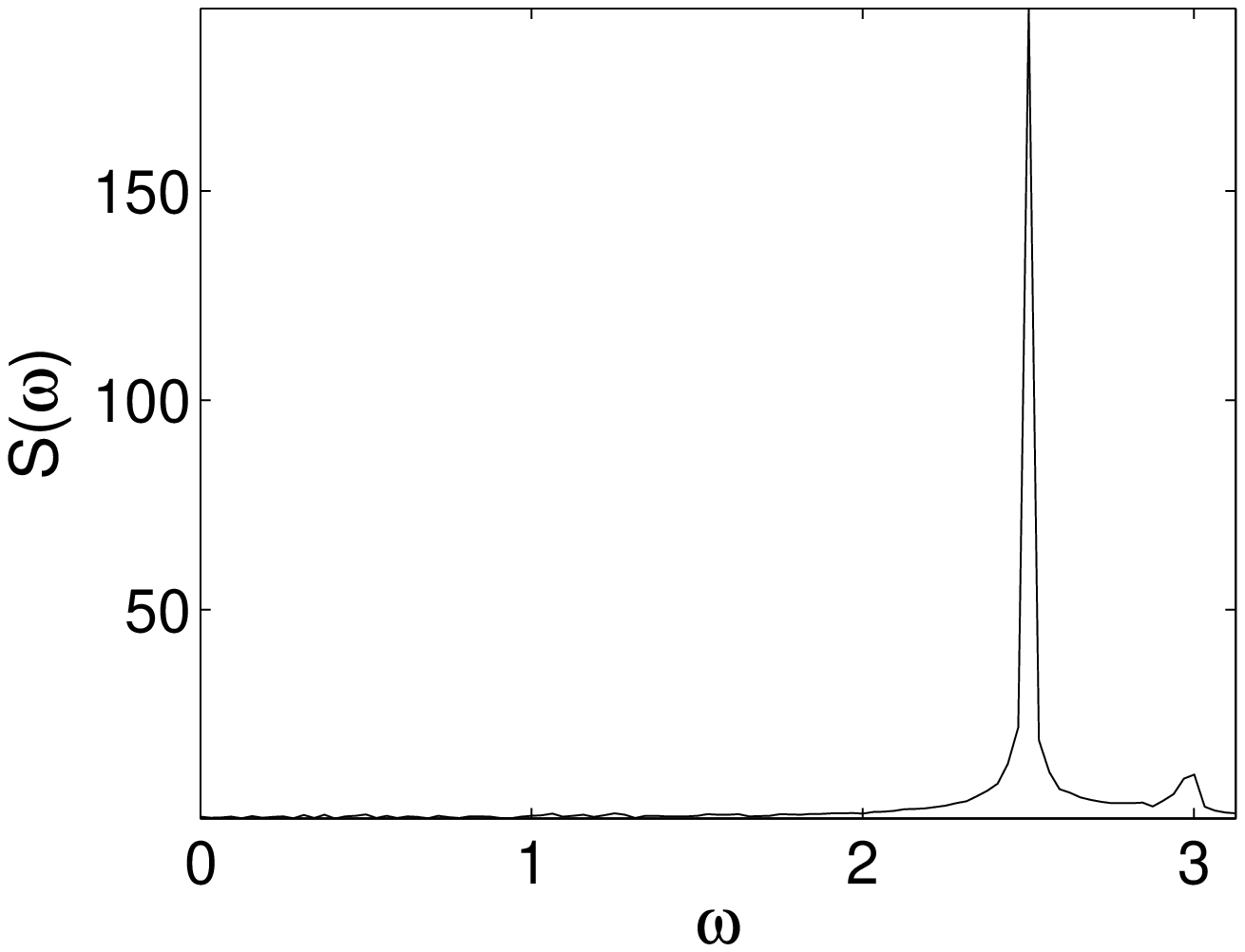}
\end{center}
\caption{Trapping: Contour plot corresponding to the power of
soliton $P$ as function of site $n$ and time $t$ (top panel) and
Fourier components of the trapped soliton calculated soon after
the collision (bottom panel). The parameters are $\alpha=0.2$, $q=0.3$,
$\omega=2.5$, $C=1$ and the impurity is located at $n=0$.}
\label{Fourier}
\end{figure}

\item[(b)] Trapping and reflection. If the impurity is attractive,
but strong
enough, some fraction of energy remains trapped by the impurity, but
a considerable amount of it is reflected. The reflected excitation
remains localized. This case is similar to the previous one, but now the
incident traveling structure has enough energy and norm to excite a
stationary mode centered at the impurity, remaining localized
and give rise to a reflected pulse. A
typical case is shown in Fig.~\ref{Fourier2}, corresponding to
$q=0.6$ and $\alpha=1.0$. The incident wave has power and energy
$P=2.61$ and $E=-4.79$, and the stationary nonlinear mode centered
at the impurity, with the same frequency, $P=0.76$ and $E=-1.79$.
When the incident breather reaches the impurity, it excites the
nonlinear mode, and, after losing some energy (in absolute value),
part of it remains localized, and another part is reflected. Also, in
our numerical simulations, we have detected, as in the previous
case, that the frequency of the remaining trapped mode is slightly
lower than that the incident breather, so it has even smaller energy (in
absolute value) and power than the corresponding nonlinear mode with
the frequency of incident soliton.

\begin{figure}
\begin{center}
\includegraphics[scale=0.5]{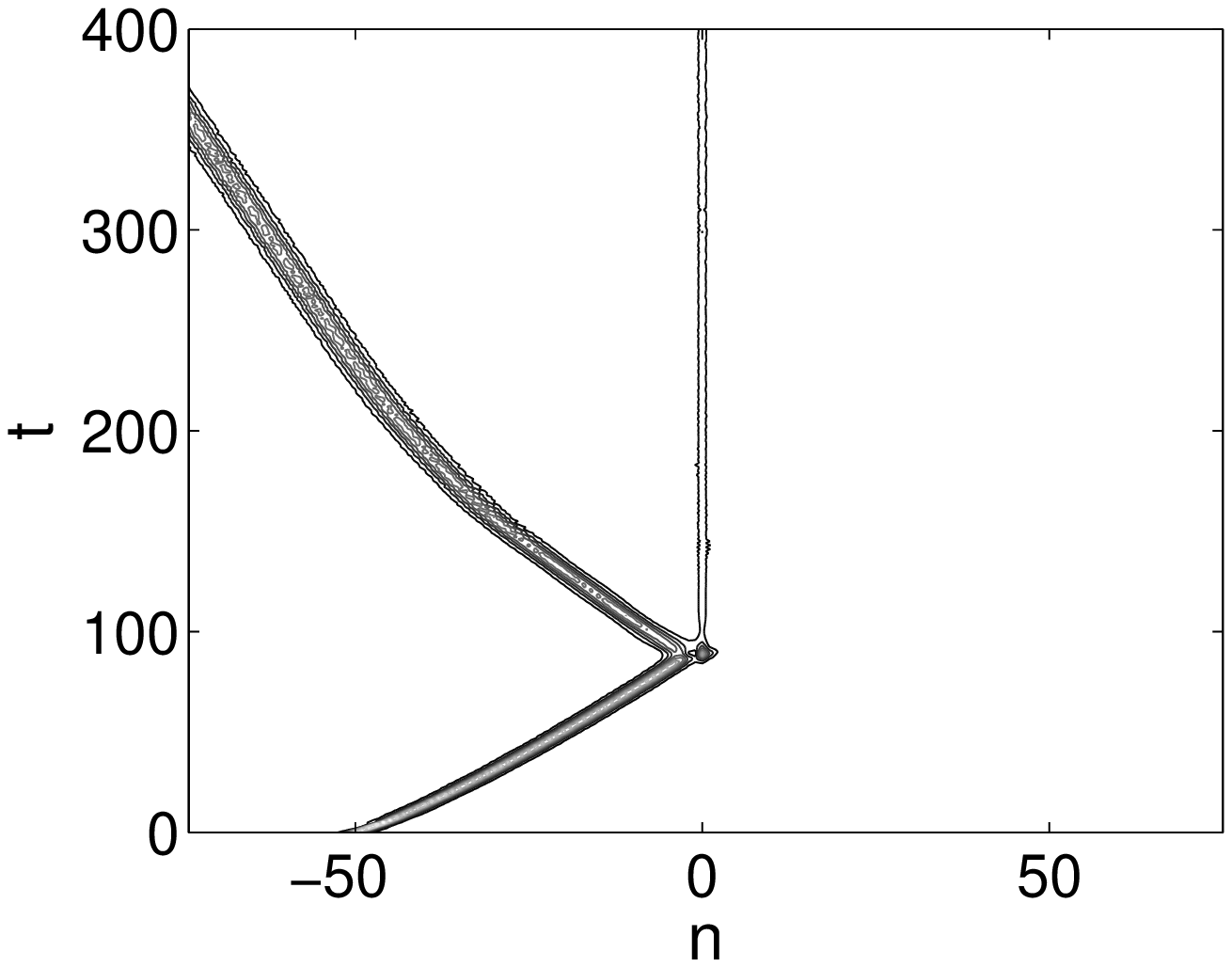}\\[1.0ex]
\includegraphics[scale=0.5]{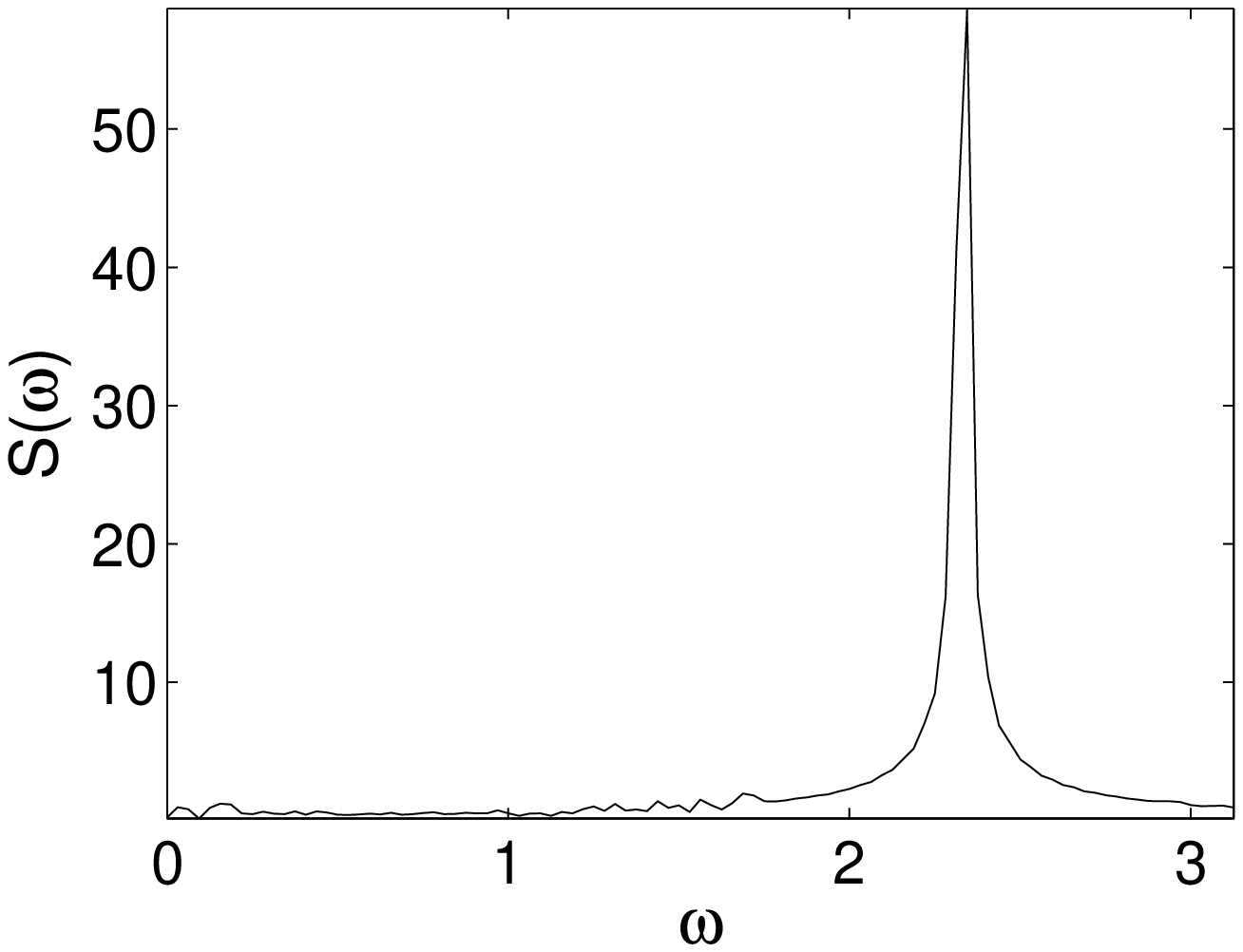}
\end{center}
\caption{Trapping and reflection:  Contour plot corresponding to
the power of soliton $P$ as a function of site $n$ and time $t$
(top panel) and Fourier components of the trapped soliton calculated
soon after the collision (bottom panel). The parameters are $\alpha=1.0$,
$q=0.6$, $\omega=2.5$, $C=1$ and the impurity is located at
$n=0$.}\label{Fourier2}
\end{figure}

\begin{figure}
\begin{center}
\includegraphics[scale=0.5]{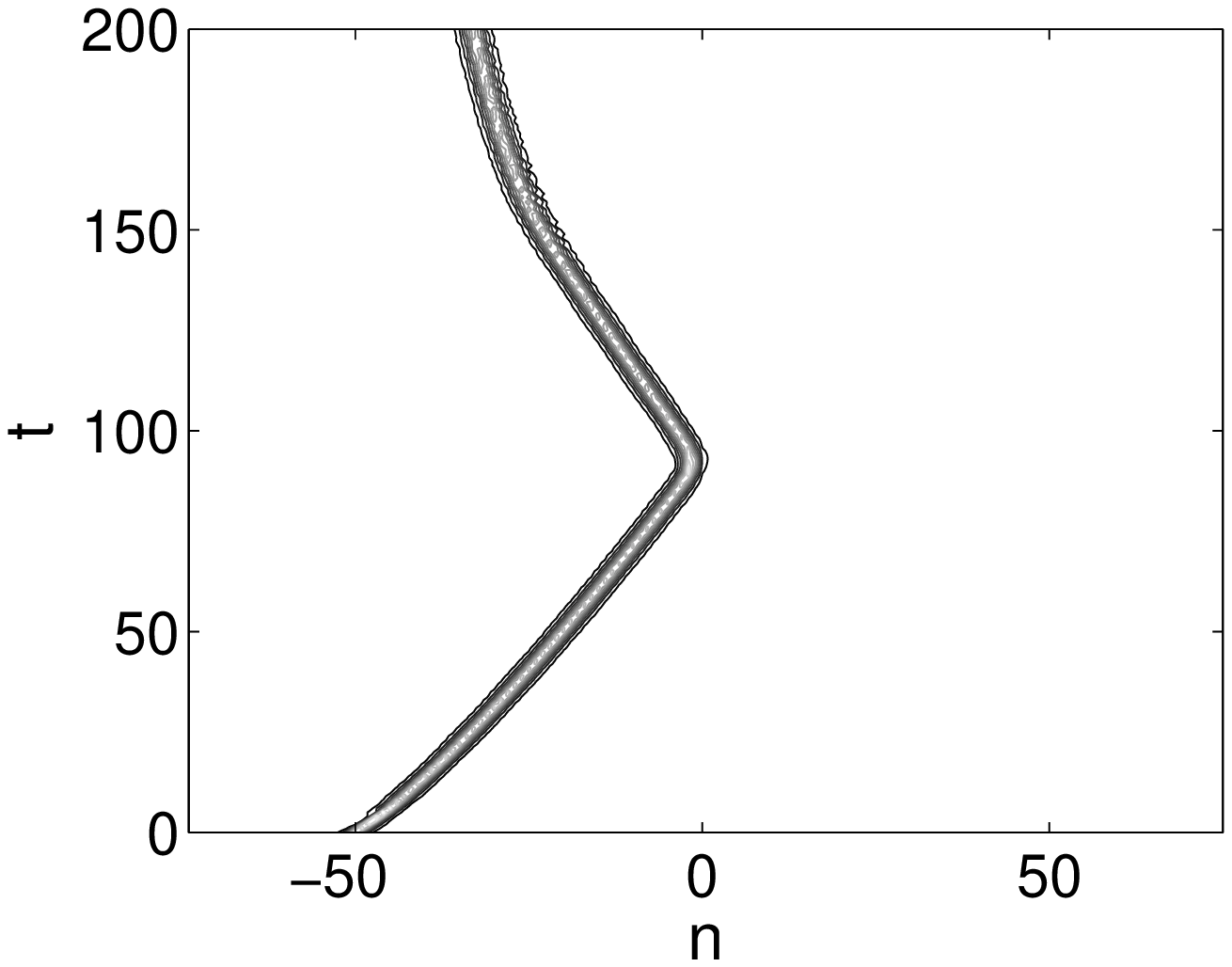}\\[1.0ex]
\includegraphics[scale=0.5]{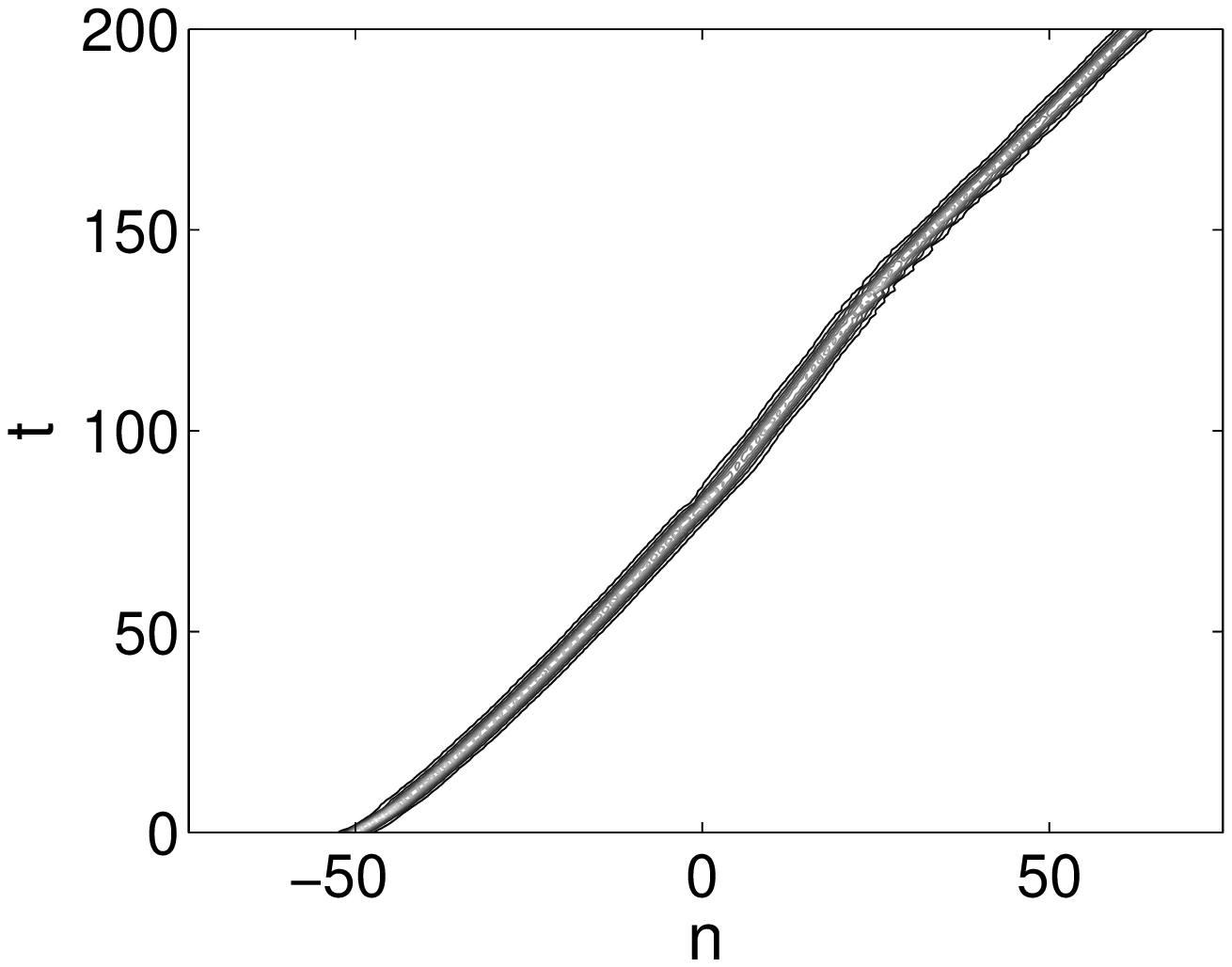}
\end{center}
\caption{Reflection with no trapping (top panel) corresponding to
parameters $\alpha=-0.5$, $q=0.6$ and $\omega=2.5$ and
transmission with no trapping (bottom panel) corresponding to parameters
$\alpha=0.1$, $q=0.7$ and $\omega=2.5$. In both cases we represent
a contour plot corresponding to the power of soliton $P$ as
function of site $n$ and time $t$, $C=1$ and the impurity is
located at $n=0$.} \label{reflex}
\end{figure}

\begin{figure}
\begin{center}
\includegraphics[scale=0.5]{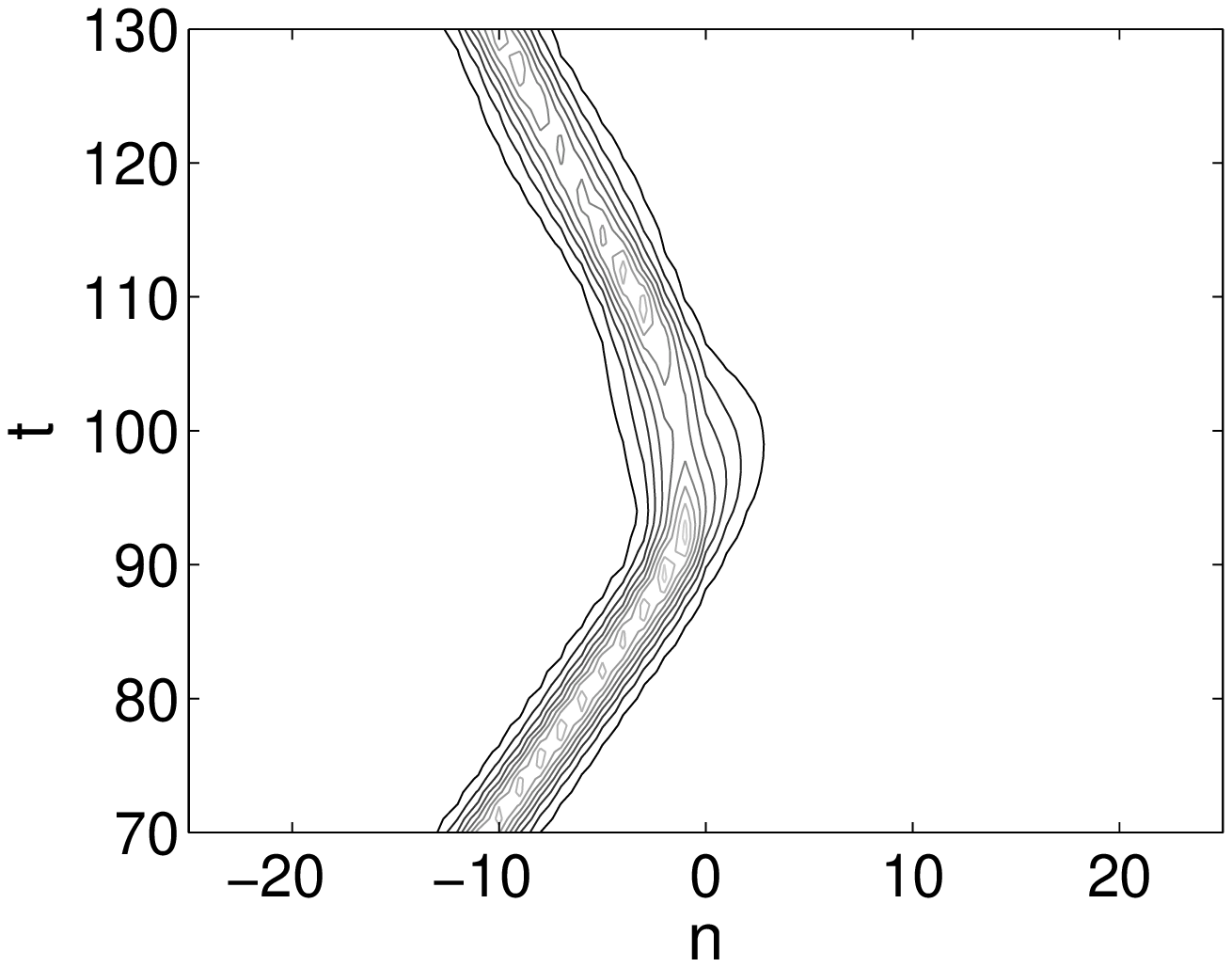}\\[1.0ex]
\includegraphics[scale=0.5]{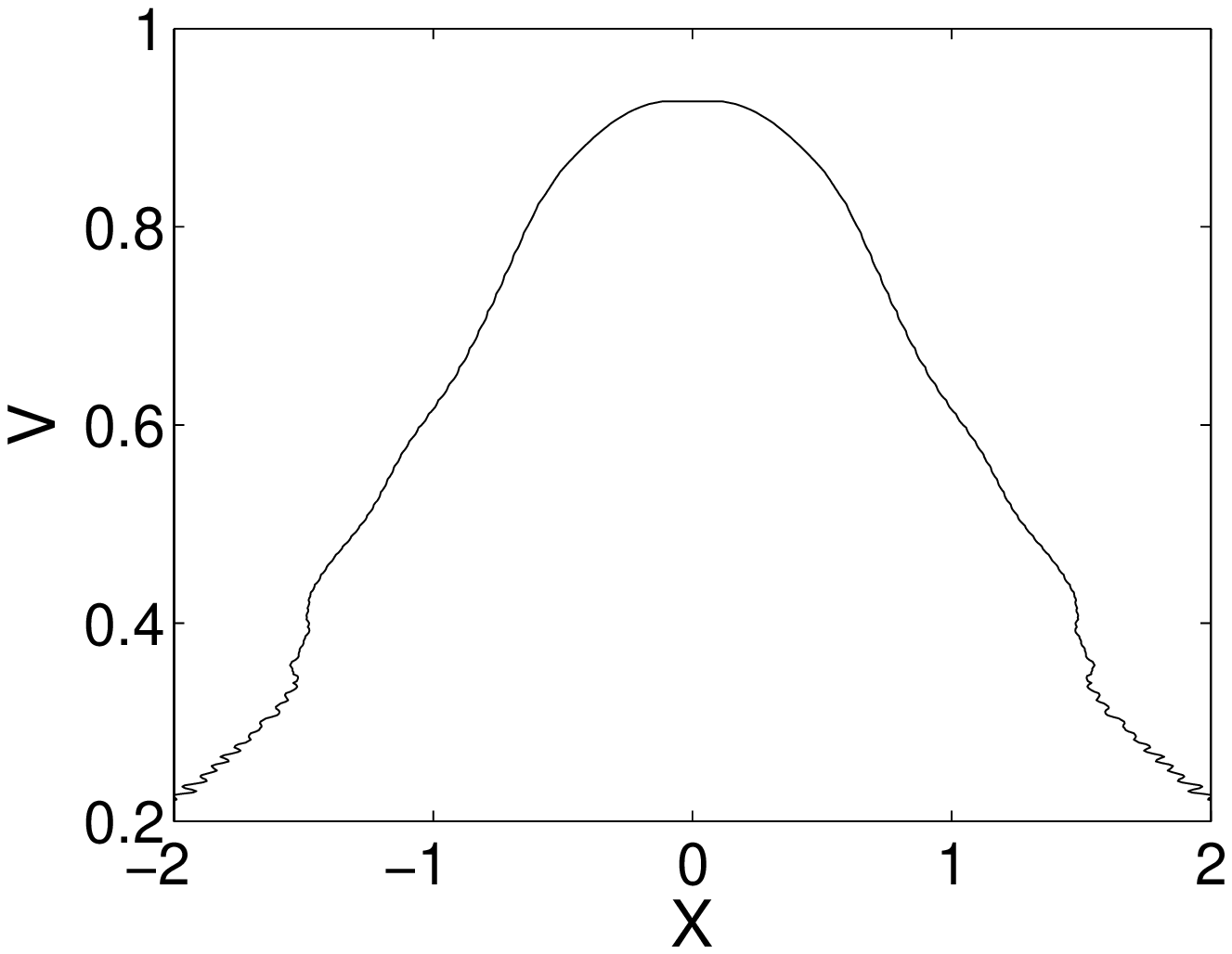}
\end{center}
\caption{Contour plot of the phenomenon of reflection of a soliton
corresponding to trust parameter $q=0.6$ (top panel). Potential barrier
calculated as described in text (bottom panel). In both cases
$\alpha=-0.2$, $C=1$, $\omega=2.5$ and the impurity is located at
$n=0$.} \label{barrier}
\end{figure}

In general, we have found that a necessary condition to trap
energy and power by the impurity is the existence of a nonlinear
localized mode centered at the impurity, with similar frequency,
and energy (in absolute value) and power smaller than that of the
corresponding incident soliton.

\item[(c)] Reflection with no trapping. Here,
we have to distinguish two cases. If the impurity is repulsive,
and $q$ small enough, neither trapping, nor transmission occur.
Instead, all energy is reflected, and the traveling nonlinear
excitation remains localized. In this case, as shown in
Fig.~\ref{reflex} (top), the incident wave has no energy and power to
excite the localized mode. In a typical case, i.e., $\omega=2.5$,
$q=0.6$ and $\alpha=-0.5$, the incident soliton has energy and
power $E=-4.79$ and $P=2.61$, and the nonlinear localized mode on
the impurity with the same frequency $E=-8.038$ and $P=3.77$. No
trapping phenomenon occurs, and the pulse is reflected.

On the other hand, if the impurity is attractive and strong enough, i.e.
$q=0.7$, $\omega=2.5$ and $\alpha=2.0$, the frequency of the soliton
is smaller than the corresponding  to linear impurity mode
($\omega_{L}\simeq 2.82$), and all the energy is reflected. This is in
accordance with the necessity of a nonlinear localized mode at the
impurity site in order for the trapping to occur.

\item[(d)] Transmission with no trapping. If $|\alpha|$ is small enough,
and $q$ high enough, transmission with no trapping occurs, as
shown in Fig.~\ref{reflex} (bottom). There exists a critical value
of $q=q_c>0$ that, if $q>q_c$, the incident soliton crosses
through the impurity. The value of $q_c$ grows with $|\alpha|$. In
the case where $q<q_c$, if $\alpha<0$, reflection with no trapping
occurs, while if $\alpha>0$, trapping with no reflection
phenomenon takes place.
\end{itemize}

Our results related to trapping, reflection and transmission
phenomena are in agreement with some results recently obtained,
using a different approach, in a similar system \cite{molina}. In
this work, where approximate discrete moving solitons with fixed
amplitude are generated using a continuous approximation, the
authors study the trapping process by a linear and a nonlinear
attractive impurity. In this latter framework, trapping can be explained
by means of resonances with the linear localized mode. In our
case, where nonlinear effects become stronger, all this
phenomena are related with resonances with a nonlinear localized
mode.

Finally, a very interesting phenomenon occurs when parameter $\alpha$ is repulsive and small (in
absolute value) enough. In this case, the solitary wave can be reflected or transmitted depending
on its velocity. Also, when it is reflected,  our numerical tests show that its velocity is similar
to its incident velocity. Thus, if we consider the soliton as a ``quasiparticle'', the effect of
the impurity is similar to the effect of a potential barrier. To determine this potential barrier
for a given value of parameter $\alpha$, we have used a method similar to the one described by Ref.
\cite{Cue04}. We have considered different values of the thrust parameter $q$ corresponding to the
reflection regime, and determine, for each value, the turning point, $X(q)$. Thus the translational
energy of the barrier for this value of $q$ is defined as the difference between the energy
(\ref{ham}) of the moving soliton (\ref{thrust}) and the stationary state (\ref{stat}) of the same
frequency far from the impurity. It can be written as $V(q)=C\sin(q/2)|P(q/2)|$, with
$P(q)=i\sum_n\psi_n^*\psi_{n+1}-\psi_n^*\psi_{n-1}$ being the lattice momentum, as defined in Ref.
\cite{Papa03}. Results are shown in Fig.~\ref{barrier}, which exhibits, as expected, an irregular
shape, whose origin lies in the nonuniform behavior of the translational velocity due to the
discreteness of the system.

On the other hand, if the parameter $\alpha$ is small enough, and positive (attractive), the
solitary wave faces a potential ``well'' and can be trapped if its translational energy is small
or, if the translational energy is high enough, it may be transmitted, losing energy that remains
trapped by the impurity, and decreasing its velocity. We have not found a regime with trapping and
transmission as have also been observed in Klein--Gordon lattices \cite{Cue02}.

\section{Conclusions}

In this work, we have revisited the long-standing theme
of the interactions of DNLS localized modes with an impurity.
In particular, we have examined both the case of attractive
and repulsive impurities and have shown how localized
modes bifurcate out of the linear spectrum in the presence
of the impurity. Subsequently, we have seen how
{\em drastically} the bifurcation
diagram of localized modes is affected by the presence of
the impurity. In particular, we have concluded that for
attractive impurities the on-site mode at the impurity eventually
disappears, while for repulsive ones, it becomes unstable
beyond a critical impurity strength. In addition, localized
modes one site away from the impurity and beyond are also
structurally affected and cannot be sustained under strong
(either attractive or repulsive) impurities. Furthermore,
we have seen how the presence of the impurity significantly
modifies the threshold for the formation of localized modes,
under a compactum of initial data. Attractive impurities
favor the formation of such a mode under weaker excitations,
while repulsive ones necessitate an even higher amplitude threshold.
Finally, we have examined in detail for both impurity cases
(attractive and repulsive) the interaction of the impurity
with a moving localized mode initiated away from it. The
principal regimes that we have identified as a function
of the impurity strength (and sign) and initial speed
are trapping, partial trapping and partial reflection,
pure reflection and pure transmission. In general,
if the impurity is repulsive, and the speed small enough,
the wave is always
reflected. If the impurity strength
(in absolute value) is small enough and the speed
is high enough, then transmission can take place.
On the other hand, if
impurity is attractive, trapping can occur, and if the speed is high
enough trapping with reflection too. If impurity is attractive and
sufficiently strong, the frequency of the soliton is smaller than the one
corresponding to the linear localized impurity mode and the wave is
reflected.

There are numerous avenues that one can think of for further
exploration of this subject. On the one hand, we feel that numerous
among the conclusions of the present work including ones about
the unavailability of localization on or at nearby sites
to the impurity for sufficiently high strengths, or ones about
the threshold for localized modes should be immediately experimentally
testable in arrays of optical waveguides. On the other hand,
this type of wave-impurity interactions have been predominantly
studied in one-dimensional systems. However, the present availability
of two-dimensional waveguide arrays renders this a very interesting
system for examining the relevant interaction in multi-dimensional
frameworks, even from a theoretical point of view and the examination
of both the standing wave and of the scattering problems.
The latter problem is currently under investigation and will
be reported in future publications.

\section*{Acknowledgments}
FP and JC acknowledge financial support from the MECD project
FIS2004-01183. FP thanks San Diego State University for
hospitality, and the Secretar\'{\i}a de Estado de Universidades e
Investigaci\'on del Ministerio de Educaci\'on y Ciencia (Spain)
for financial support. RCG and PGK gratefully acknowledge support
from NSF-DMS-0505663. PGK also acknowledges support from
NSF-DMS-0619492 and NSF-CAREER.

\appendix

\section{Invariant manifolds approximation}

In this appendix we sketch the method followed in Section 4.1.4 of
Ref.~\cite{Gui07} for determining the value of $\alpha_c$, i.e., the
value of $\alpha$ at which the breathers centred at $n=n_0$ and
$n=n_0+0.5$ bifurcate.

The difference equation (\ref{SDNLS}), for $\alpha=0$, can be recast as a
two-dimensional real map by defining $y_n=\phi_n$ and
$x_n=\phi_{n-1}$ \cite{DNLSmap}:
\begin{equation}\label{map}
    \left\{\begin{array}{l}
    x_{n+1}=y_n \\[2.0ex]
    y_{n+1}=(\omega y_n-y_n^3)/C-x_n .
    \end{array}\right.
\end{equation}

For $\omega>2$, the origin $x_n=y_n=0$ is
hyperbolic and
a saddle
point. Consequently, there exists a one-dimensional stable ($W^s(0)$)
and a one-dimensional
unstable ($W^u(0)$) manifolds emanating from the origin in two
directions given by $y=\lambda_{\pm}x$, with
\begin{equation}\label{eigen}
    \lambda_{\pm}=\frac{\omega\pm\sqrt{\omega^2-4C^2}}{2C} .
\end{equation}

\begin{figure}
\begin{center}
\includegraphics[scale=0.4]{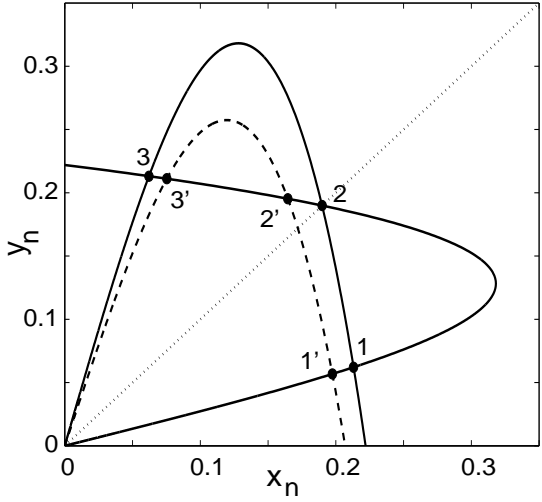}
\end{center}
\caption{First winding of the homoclinic tangle of the map
(\ref{map}). Dashed line corresponds to the linear transformed
unstable manifold when $\alpha=0$. Labels $1$, $2$, $3$ ($1'$,
$2'$, $3'$) corresponds to fundamental solitons for $\alpha=0$
($\alpha\neq0$).} \label{tangle}
\end{figure}

These manifolds intersect in general transversally, yielding the
existence of an infinity of homoclinic orbits. Each of their
intersections corresponds to a localized solution. Fundamental
solitons (i.e. on-site and inter-site solitons), correspond to the
primary intersections points, i.e. those emanating from the first
homoclinic windings. Each intersection point defines an initial
condition $(x_0,y_0)$, that is, $(\phi_{-1},\phi_0)$, and the rest of the
points composing the soliton are determined by application of the
map (\ref{map}) and its inverse.
Fig.~\ref{tangle} shows an example of the first windings of the
manifolds. Intersections corresponding to fundamental solitons are
labeled as follows: (1) is the on-site breather centred at $n=0$,
(2) is the inter-site breather centred at $n=0.5$ and (3) is the
on-site breather centred at $n=1$.

The effect of the inhomogeneity is introduced as a linear
transformation of the unstable manifold $A(\alpha)W^u(0)$ with
$A(\alpha)$ given by:

\begin{equation}
\label{defA} A(\alpha)= \left(
\begin{array}{cc}
1 & 0  \\[1.0ex]
-\alpha/C & 1
\end{array}
\right)
\end{equation}

When $\alpha>0$, the unstable manifold moves downwards, changing the
intersections between the transformed unstable manifold and the
stable manifold to points $1'$, $2'$ and $3'$ (see Fig.~\ref{tangle}).
For $\alpha=\alpha_c$, both manifolds become tangent.
Thus, for $\alpha>\alpha_c$ intersections $3'$ and $2'$ are lost,
that is, for $\alpha=\alpha_c$ the breathers centred at $n=1$ and
$n=0.5$ experience a tangent bifurcation. On the contrary, if
$\alpha<0$, intersections $1'$ and $2'$ are lost when
$|\alpha|>|\alpha_c|$, leading to a bifurcation between the
breathers centered at $n=0.5$ and $n=0$.

A method for estimating $\alpha_c(\omega)$ is based on a simple
approximation of $W^u(0)$. Let us consider a cubic approximation
$W^u_{\rm app}$ of the local unstable manifold of Fig.~\ref{tangle},
parametrized by $y = \lambda\, x - c^2 \, x^3$, with
$\lambda\equiv\lambda_+$. The coefficient $c$ depends on $\omega$
and $C$ and need not be specified in what follows (a value of $c$
suitable when $\lambda$ is large is computed in Ref.~\cite{Cue07}). We
have
\begin{equation}
\label{awu} y = \lambda_0 x - c^2 \, x^3
\end{equation}
on the curve $ A(\omega ,\alpha)W^u_{\rm app}$, where $\lambda_0=\lambda
- \alpha/C $. By symmetry we can approximate the local stable
manifold using the curve $W^s_{\rm app}$ parametrized by
\begin{equation}
\label{ws} x = \lambda\, y - c^2 \, y^3 .
\end{equation}
The curves $ A(\alpha)W^u_{\rm app}$ and $W^s_{\rm app}$ become tangent at
$(x , y )$ when in addition
\begin{equation}
\label{tan} (\lambda  - 3 c^2 \, x^2 ) (\lambda_0  - 3 c^2 \, y^2
)=1.
\end{equation}

In order to compute $\alpha_c$ as a function of $\omega$, or,
equivalently, the corresponding value of $\lambda_0$ as a function
of $\lambda$, one has to solve the nonlinear system
(\ref{awu})--%
(\ref{tan}) with respect to $x$, $y$,
$\lambda_0$, which yields a solution depending on $\lambda$. Instead
of using $\lambda$ it is practical to parametrize the solutions by
$t=y / x$. This yields
$$
x = \frac{1}{c\sqrt{2}}\, (t+\frac{1}{t^3})^{1/2}, \ \ \ y =
\frac{t}{c\sqrt{2}}\, (t+\frac{1}{t^3})^{1/2},
$$
$$
\lambda_0 = \frac{3}{2} t + \frac{1}{2 t^3}, \ \ \ \lambda =
\frac{3}{2 t}  + \frac{1}{2}\, {t^3}.
$$
Since $\lambda + \lambda^{-1}=\omega/C$  it follows that
\begin{equation}
\label{eqmut} t^4-2\lambda t+3 = 0,
\end{equation}
\begin{equation}
\label{mcapp} \alpha = \frac{C}{2} (t-\frac{1}{t})^3.
\end{equation}
Given a value of $\omega$, one can approximate $\alpha_c$ by the
value of $\alpha$ given by equations (\ref{eqmut})--(\ref{mcapp}). In
particular, (\ref{eqmut}) has two real positive solutions (one
larger than 1, and another smaller than 1), and two complex
conjugated solutions. The solution with $t>1$ ($t<1$) leads to $\alpha_c>0$
($\alpha_c<0$) and, subsequently, approximates the tangent
bifurcation values when the breathers at $n=0.5$ and $n=1$ ($n=0$)
collides.

Despite it gives precise numerical results in a certain parameter
range, the approximation (\ref{eqmut})--(\ref{mcapp}) is not always
valid. Indeed, the parameter regime $\omega < 5C/2$ is not described
within this approximation \cite{Cue07}.

\end{document}